\pdfoutput=1

\documentclass[preliminary,creativecommons]{eptcs}
\providecommand{\event}{arXiv: \href{https://arxiv.org/abs/2012.02154}{2012.02154}}
\usepackage[utf8]{inputenc}

\usepackage{underscore}

\usepackage[T1]{fontenc}
\usepackage{mathpartir}
\usepackage{amssymb}
\usepackage{amsmath}
\usepackage{amsthm}
\usepackage{braket}
\usepackage{xspace}
\usepackage{microtype}
\usepackage[capitalize,nameinlink]{cleveref}
\crefformat{section}{#2\S#1#3}
\crefformat{lstlisting}{#2Listing~#1#3}
\crefmultiformat{lstlisting}{Listings~#2#1#3}%
{ and~#2#1#3}{, #2#1#3}{ and~#2#1#3}
\crefrangeformat{lstlisting}{Listings~#3#1#4--#5#2#6}

\usepackage{tikz}
\usetikzlibrary{quantikz}

\newcommand{\mailtodomain}[1]{\href{mailto:#1}{\nolinkurl{#1}}}

\usepackage{xcolor}
\definecolor{dkgreen}{rgb}{0,0.35,0}
\definecolor[named]{ACMPurple}{RGB}{101,1,107}
\definecolor[named]{ACMDarkBlue}{RGB}{9,53,122}

\usepackage{listings}
\lstdefinelanguage{QHaskell}{
    language     = Haskell,
    morekeywords = {bit, qbit, vector, complex, unit, meas, apply, to, requires, ensures, unitary},
}

\newcommand{\HoareT}[3]{
    \{#1\} ~#2~ \{#3\}
}

\theoremstyle{definition}
\newtheorem{definition}{Definition}[section]
\theoremstyle{remark}
\newtheorem*{remark}{Remark}

\usepackage{ottalt}

\newenvironment{ottdefnblock}[3][]{ \framebox{\mbox{#2}} \quad #3 \\[0pt]}{}

\newcommand{\ottafterlastrule}{\\}

  \renewottcommands[ott]

\usepackage{hyperxmp}
\def\titlerunning{Quantum Hoare Type Theory}
\def\authorrunning{Kartik Singhal}
\hypersetup{
    colorlinks,
    linktoc=all,
    linkcolor=ACMPurple,
    citecolor=ACMPurple,
    urlcolor=ACMDarkBlue,
    filecolor=ACMDarkBlue,
    pdftitle={\titlerunning},
    pdfauthor={\authorrunning},
    pdfkeywords={General programming languages, Quantum computation, Functional languages, Program specifications, Pre- and post-conditions, Type theory, Hoare logic},
    pdfsubject={Computer Science Master's paper},
    pdfapart=2,
    pdfaconformance=B
}
\immediate\pdfobj stream attr{/N 3} file{sRGB.icc}
\pdfcatalog{%
  /OutputIntents [
    <<
      /Type /OutputIntent
      /S /GTS_PDFA1
      /DestOutputProfile \the\pdflastobj\space 0 R
      /OutputConditionIdentifier (sRGB)
      /Info (sRGB)
    >>
  ]
}

\usepackage[figure,table]{hypcap}

\usepackage[style=ext-authoryear-comp,dashed=false,maxbibnames=99,isbn=false,backref=true]{biblatex}

\DeclareFieldFormat{citehyperref}{%
  \DeclareFieldAlias{bibhyperref}{noformat}%
  \bibhyperref{#1}}

\DeclareFieldFormat{textcitehyperref}{%
  \DeclareFieldAlias{bibhyperref}{noformat}%
  \bibhyperref{%
    #1%
    \ifbool{cbx:parens}
      {\bibcloseparen\global\boolfalse{cbx:parens}}
      {}}}

\savebibmacro{cite}
\savebibmacro{textcite}

\renewbibmacro*{cite}{%
  \printtext[citehyperref]{%
    \restorebibmacro{cite}%
    \usebibmacro{cite}}}

\renewbibmacro*{textcite}{%
  \ifboolexpr{
    ( not test {\iffieldundef{prenote}} and
      test {\ifnumequal{\value{citecount}}{1}} )
    or
    ( not test {\iffieldundef{postnote}} and
      test {\ifnumequal{\value{citecount}}{\value{citetotal}}} )
  }
    {\DeclareFieldAlias{textcitehyperref}{noformat}}
    {}%
  \printtext[textcitehyperref]{%
    \restorebibmacro{textcite}%
    \usebibmacro{textcite}}}

\DeclareNameAlias{sortname}{given-family}
\DeclareOuterCiteDelims{textcite}{}{}
\DeclareInnerCiteDelims{textcite}{\bibopenbracket}{\bibclosebracket}
\DeclareOuterCiteDelims{parencite}{\bibopenbracket}{\bibclosebracket}
\DeclareInnerCiteDelims{parencite}{}{}
\DeclareUnicodeCharacter{0301}{\'{\i}}

\title{\titlerunning}

\author{
\authorrunning
\institute{University of Chicago}
\email{\mailtodomain{ks@cs.uchicago.edu}}
}

\begin{document}

\maketitle

\begin{abstract}
    As quantum computers become real, it is high time we come up with effective techniques that help programmers write correct quantum programs. Inspired by Hoare Type Theory in classical computing, we propose Quantum Hoare Type Theory (QHTT), in which precise specifications about the modification to the quantum state can be provided within the type of computation. These specifications within a Hoare type are given in the form of Hoare-logic style pre- and postconditions following the propositions-as-types principle. The type-checking process verifies that the implementation conforms to the provided specification. QHTT has the potential to be a unified system for programming, specifying, and reasoning about quantum programs.
\end{abstract}

\thispagestyle{empty}

\tableofcontents

\listoftables

\listoffigures

\lstlistoflistings

\section{Introduction}

Quantum computation is fast becoming a reality with the advent of real machines and Google's recent breakthrough~\parencite{supremacy19} in demonstrating so-called \textit{quantum supremacy}.\footnote{A term coined by \textcite{Preskill2012} to describe the moment when a quantum computer could perform a computational task that no classical computer could. Alternatives include \textit{quantum advantage}~\parencite{nature-advantage2019} and \textit{quantum primacy}~\parencite{durham2021}.} With advances in hardware, there is an acceleration in the development of software for quantum machines, which in turn requires the development of special-purpose programming languages. Several efforts in this direction, both from academia and industry, have recently led to many programming languages aimed at quantum computing such as Q\#~\parencite{qsharp2018}, Quipper~\parencite{quipper2013}, Scaffold~\parencite{scaffold12}, and Silq~\parencite{silq20}.

There is also a sudden need to prepare and train the upcoming generation of quantum programmers who sometimes have little to no previous exposure to the fundamental concepts in quantum computation. If we were to put ourselves in the position of a beginner to quantum programming, we might start by learning basic concepts such as what is a qubit, what does it mean for a qubit to be in a superposition of 0 and 1, and the most perplexing of all: entanglement. We may then try to write some simple programs in a suitable quantum programming language to aid our understanding. We may even be able to run some of these programs on simulators and even on real machines. We will make mistakes when writing these programs, however, and while sometimes those mistakes may be apparent from the output, a lot of times (as a beginner), they may be not. What can we do?

A common method to test our assumptions while programming in the classical realm is to write assertions about what we believe to be the truth at a particular point in the program. These assertions are usually dynamic in nature, meaning they are only tested while running the program and involve a cycle of writing, running, debugging, and rerunning the program. While this may be a suitable approach in classical computing, it is quite wasteful of the limited quantum resources if we were to apply it to quantum programming. Furthermore, this approach is impossible to use if the number of qubits is just over 50 or so as we hit the limits of both current quantum machines and classical simulation. Despite these limitations, there have been many recent proposals to use run-time assertions for debugging quantum programs \parencite{Li2020,huang2019}.

What can we from the programming languages community offer that can help the upcoming workforce of quantum programmers avoid mistakes in their programs?

In classical software development, the use of strong static type systems has been proven to be hugely beneficial in avoiding large classes of bugs before running the programs. Recent mainstream languages such as Rust~\parencite{matsakis2014} even manage to statically prevent memory safety bugs~\parencite{xu2021memorysafety,Jung2017,Jung2021} that have been notorious for being the root causes of a large swath of security vulnerabilities found in production software written in unsafe languages such as C. There is a similar need for innovation in type systems for quantum programming languages.

Recent academic languages, such as QWire~\parencite{qwire2017,paykin2018} and Silq~\parencite{silq20}, that employ a linear type system to statically enforce the no-cloning theorem of quantum mechanics show promise in the approach of statically preventing bugs in the quantum domain. However, semantic properties about the quantum state, such as whether a qubit is in superposition or whether it is entangled with some other qubit, have not been tackled with the help of types. Furthermore, assertions in the form of pre- and postconditions around small pieces of code have shown the most promise in the quantum domain \parencite{huang2019,qdb2018}. This seems fair: we, as humans, can only hold so much information about the quantum state in our brains, it makes sense to reason about pieces of quantum code in a modular manner.

The most common approach to reason with pre- and postconditions in the classical domain is to use variations of Floyd-Hoare logic that utilize specification triples of the form $\HoareT{P}{c}{Q}$, where $c$ is the program to be executed, $P$ a proposition on the initial state associated with the program (precondition), and $Q$ that on the final state (postcondition). The intuitive reading of such a triple is that if the precondition holds on the program state before executing the program, and the program terminates, then the postcondition must also hold. These triples, along with a set of axioms and inference rules for composing them together, have proven to be quite effective in proving the correctness of classical state-manipulating (imperative) programs over the last several decades. It is no surprise that similar logics have appeared in the quantum computing domain as well \parencite{aqhl2019,floydhoare2012,unruh2019,wpe2016}.

Hoare Type Theory (HTT)~\parencite{nanevski2008} encodes Hoare-style specifications as types that can then be used to enforce desired properties about the program state statically. This encoding is made possible with the use of dependent types that can depend on values instead of just other types and which add immense expressive power to a programming language. In effect, we obtain a functional programming language in which imperative code is encapsulated within a computation type which is indexed by the type of the result of the computation and its specification in the form of pre- and postconditions. We then obtain a system where the process of type checking a program becomes equivalent to verifying that the program meets the given specification. In this work, we consider the problem of reasoning about quantum programs in HTT and present our Quantum Hoare Type Theory (QHTT).

To obtain a quantum variant of HTT, we replace the classical primitive commands in its original formulation with quantum-specific primitive commands and, taking inspiration from a novel approach by \textcite{unruh2019}, employ ghost variables for reasoning locally about the quantum state in our quantum proposition logic. Further, the use of ghost variables opens the door to an expressive language of propositions in which we can specify predicates about the equality, superposition, and entanglement of qubits.

Our approach takes inspiration from previous foundational work in reasoning about quantum programs such as \citetitle{dhondt2006}~\parencite{dhondt2006} and \citetitle{floydhoare2012}~\parencite{floydhoare2012} in the spirit of \textcite{hoare1969}, \textcite{dijkstra1976}, and \textcite{gries1981}, but attempts to bring those reasoning techniques into a type system. The hope is that programmers will be able to encode some semantic properties that they expect of their programs as specifications in their code, and that type checking will ensure the correctness of their implementation to those specifications. In the classical setting, Hoare Type Theory accomplishes exactly this goal. The work presented here is an attempt to merge these ideas for the emerging landscape of quantum computing.

We limit our presentation to quantum programs that do not involve iteration. This is not to suggest that iteration is not supported in the system, only that we are yet to study it.

Our investigation suggests that QHTT could be a unified system for programming, specifying, and reasoning about quantum programs.

\section{Background}
In this section, we provide background on the core ideas from the existing literature that form the foundation for our work---Hoare logic, Hoare Type Theory, quantum computation, and quantum Hoare logic with ghost variables.

\subsection{Floyd--Hoare logic}
In the late 1960s, Robert Floyd and Tony Hoare came up with a set of axioms and rules of inference that can be used to prove the properties of computer programs. This approach is commonly just known as Hoare logic since Hoare was the first to apply it to text-based programs~\parencite{hoare1969}, while Floyd had applied it to flowcharts~\parencite{floyd67}. Here we present the axioms and rules of Hoare logic in a natural deduction style.

As noted above, this technique involves the use of \textit{Hoare triples} of the form:
\begin{mathpar}
    \HoareT{P}{c}{Q}
\end{mathpar}
which can be interpreted as ``if the precondition $P$ is true before the execution of the program $c$, then the postcondition $Q$ will be true on its completion.''

In the classical setting, the most important command is an assignment statement. Hoare logic states the following axiom for assignment:
\begin{mathpar}
    \inferrule[Axiom of Assignment]
    {}
    {\HoareT{P[x \rightarrow e]}{x := e}{P}}
\end{mathpar}
where $x$ is an assignable (variable) and $e$ is a pure expression. The meaning of this axiom is that the precondition is obtained from the postcondition $P$ by substituting $e$ for all free occurrences of $x$ in $P$.

Furthermore, there are rules of inference that are required to derive new theorems from one or more axioms or theorems already proved. We first look at the two \textit{rules of consequence}. In the following notation, propositions above the line are called \textit{premises} and the proposition below the line is called \textit{conclusion.}
\begin{mathpar}
    \inferrule[Weakening the postcondition]
    {\HoareT{P}{c}{Q} \\ Q \Rightarrow R}
    {\HoareT{P}{c}{R}}

    \inferrule[Strengthening the precondition]
    {P \Rightarrow Q \\ \HoareT{Q}{c}{R}}
    {\HoareT{P}{c}{R}}
\end{mathpar}

The first rule says that if the execution of a program $c$ ensures the truth of postcondition $Q$, then it also ensures the truth of any proposition $R$ logically implied by $Q$. This effectively lets us weaken the postcondition. In the other direction, we can strengthen the precondition using the second rule, which says that if $Q$ is a precondition of program $c$, then so is any other proposition $P$ that logically implies $Q$.

Alternatively, we can provide a single rule combining the two above:
\begin{mathpar}
    \inferrule[Rule of consequence]
    {P \Rightarrow Q \\ \HoareT{Q}{c}{R} \\ R \Rightarrow S}
    {\HoareT{P}{c}{S}}
\end{mathpar}

The next rule lets us compose sequential programs together:
\begin{mathpar}
    \inferrule[Rule of Composition]
    {\HoareT{P}{c}{Q} \\ \HoareT{Q}{d}{R}}
    {\HoareT{P}{c\ \textbf{;}\ d}{R}}
\end{mathpar}
where the semicolon denotes the sequential composition of programs. The rule of composition states that if the postcondition of the first program is identical to the precondition of the second program under which it satisfies its postcondition $R$, then the composition of the programs will satisfy $R$, given that the first program satisfies its precondition $P$.

These are the most important rules that we need to understand the formal system presented in this work. We do not present other rules such as that for iteration (which we will not need) and others that can be derived. We also note that since we do not have an iteration construct in our system, we do not run into the issue of reasoning about the termination of programs. In other words, all programs presented in our system are terminating.

There is one other crucial concept we need to understand before moving to the next section. Let us motivate it with the help of an example that will also serve the purpose of demonstrating how to use Hoare logic. Suppose we are given the following simple triple:
\begin{mathpar}
    \HoareT{x = 1}{x := 5}{x > 0}
\end{mathpar}

This obviously seems correct. But how can we mechanically use the formal system just described to show this triple is valid? The missing piece is the idea of the \textit{strongest postcondition}. Given the assignment statement, we can compute the strongest possible postcondition to be $x = 5$, which implies $x > 0$, that is, the given postcondition. Now, we can use the first rule of consequence (weakening the postcondition) to derive the given triple.

In Hoare Type Theory that we describe next and its quantum variant later, inferring the strongest postcondition for each statement will be the most important step during the process of type checking.

We note that the style of reasoning using the strongest postconditions that we described above is also called the forward reasoning style. There is a dual theory of \textit{weakest preconditions} that are computed backward starting from a postcondition. We prefer the forward style in this work as it is easier to compute the strongest postconditions automatically using symbolic execution.

\subsection{Hoare Type Theory}

A type theory is a formal system of term rewriting in which each term has a type, notated as $t:A$ for a term $t$ and a type $A$. A type theory that admits dependent types can be called a dependent type theory. Dependent type theories can be viewed as logical systems as popularized by the Curry--Howard correspondence (also known as the principle of propositions-as-types or proofs-as-programs). Under this interpretation, logical quantifiers such as $\forall$ (universal) and $\exists$ (existential) correspond respectively to the dependent function ($\Pi$) and pair ($\Sigma$) types in type theory, for example. It turns out that with the expressive power that comes with dependent types, it becomes possible to encode the behavioral specifications of programs into their types. For a given function, the output type then corresponds to a theorem with the input types serving as its hypotheses. Proving a theorem corresponds to writing a program of that type, and checking the proof translates to type checking.

Hoare Type Theory or HTT~\parencite{nanevski2008} is a specific dependent type theory that introduces a type constructor called a \textit{Hoare type} that both isolates effectful programs from the remaining functional core language and lets us specify their behavior. In its most basic form, a Hoare type is written as:
\begin{mathpar}
    \Delta.\HoareT{P}{x:A}{Q}
\end{mathpar}
where $x:A$ is the return value and its type and $P$, $Q$ are pre- and postconditions, respectively, thus incorporating the specification methodology of Hoare logic into types. An optional context, $\Delta$, for declaring \textit{ghost} (or \textit{logical}) variables may be included in the Hoare type for specifying certain variables that can be used only in the pre- and postconditions but not in the implementation of the program. Hoare types can also be seen as \textit{monads}~\parencite{moggi89} that are common in functional programming and are similarly used to isolate effects. Like other monads, such a \textit{Hoare monad} also provides a type constructor, a return operation, and a bind operation that satisfy the usual monad laws. The main difference is that the Hoare monad is not only indexed by the type of the computation but also with the pre- and postconditions that must be satisfied by the given computation. A fairly accessible introduction to Hoare Type Theory is available in the lecture notes by \textcite{perconti2012}.

How do we verify the correctness of programs in HTT? A user may provide arbitrary pre- and postconditions as the type for their program. In HTT, the process of verifying that the given triple is valid is performed during type checking. There are axioms and typing rules in HTT that correspond to the axioms for primitive commands and inference rules such as those for weakening the postcondition\footnote{Often just called the rule of consequence in the context of HTT.} and composition in Hoare logic. Each primitive effectful command in HTT requires a rule for inferring its strongest postconditions. When a user-specified postcondition needs to be checked for a given command, we need to show that it can be obtained by weakening the inferred strongest postcondition. The rule of composition is essentially driven by the monadic bind operation. The type-checking process is carefully divided into two phases. The first phase performs basic type checking and generates verification conditions by computing the strongest postconditions at each step of the program. In the second phase, the verification conditions need to be proved. The first phase is decidable, but the second phase may not be, hence for the second phase, either the verification conditions are deferred to an automated theorem prover (such as Z3~\parencite{z32008}) or need to be validated manually using an interactive theorem prover (such as Coq~\parencite{coq2020}).

Let us consider the example we saw in the previous section written in a typical syntax for HTT:\medskip\\
\indent\indent\lstinline[language=QHaskell]!x := 5 : {x = 1} r:unit {x > 0}!
\medskip\\where $r$ is the variable that holds the result of the assignment statement. Since an assignment returns nothing, we see the type of output is $unit$, and the pre- and postconditions are the same as before.

We use an alternative syntax inspired by the F* programming language~\parencite{fstar} in our presentation. F* is an evolution of HTT in that it supports multiple monadic effects and takes a hybrid approach between automated and interactive theorem proving. In F* syntax, we can write the above triple as:\medskip\\
\indent\indent\lstinline[language=QHaskell]!x := 5 : ST (r:unit) (requires {x = 1}) (ensures {x > 0})!
\medskip\\where $ST$ is the state monad that encapsulates computations that modify the state (here assignment), and it is indexed by the result of the computation and its two pre- and postconditions. The keywords $requires$ and $ensures$ provide extra clarity for people unfamiliar with Hoare logic, and explicitly naming the computational effect monad, $ST$, helps ensure multiple effects can be supported simultaneously. Inspired by this notation, we christen our quantum state monad as $QST$.

\subsection{Quantum computation}
It is well known that quantum computation can be described in terms of elementary linear algebra. We assume familiarity with basic notions such as vectors, matrices, vector spaces, inner products, bases, linear independence, etc.

Quantum computation is usually described over a specific vector space called the \textit{Hilbert space}, $\mathcal{H}$, which for our purposes can be thought of as an $m$-dimensional complex vector space ($\mathbb{C}^m$) equipped with an inner product. In this work, we often use the term ``state space'' to refer to a finite-dimensional Hilbert space.

The basic unit of information in quantum computation is a qubit that can be represented by a unit vector in the two-dimensional complex vector space, $\mathbb{C}^2$. Using Dirac's bra-ket notation, we can describe the state of an arbitrary qubit, $\ket{\psi}$, as a linear combination (or \textit{superposition}) of the computational basis vectors:
\begin{mathpar}
 \ket{\psi} = \alpha\ket{0} + \beta\ket{1}
\end{mathpar}
where
\begin{mathpar}
\ket{0}=\begin{bmatrix}
1 \\
0
\end{bmatrix} \and
\ket{1}=\begin{bmatrix}
0 \\
1
\end{bmatrix}
\end{mathpar}
and the two constants $\alpha, \beta$ are complex amplitudes that obey the \textit{normalization constraint}, $|\alpha|^2 +|\beta|^2 = 1$. This means that when measured in the computational basis, the probability of the qubit being in the classical state 0 is $|\alpha|^2$ and that for state 1 is $|\beta|^2$. Measurement is how we extract information from a quantum system, but its side effect is that it collapses the quantum state. For example, if the classical bit obtained was 1, the posterior state becomes $\ket{1}$, and we lose the superposition state. To work with multiple qubits, we need to take their tensor product, denoted by $\otimes$.\footnote{More precisely, we are taking a Kronecker product, which is a special case of the tensor product that is used to multiply vector spaces together. The $\otimes$ symbol can often be omitted, as shown in the example that follows.} The dimension, $m$, of the resulting state vector is exponential to the total number of qubits, $n$, i.e., $m = 2^n$. For example, if we take the tensor product of the basis states above, we get a four-dimensional state vector:
\begin{mathpar}
\ket{0}\otimes\ket{1} = \ket{0}\ket{1} = \ket{01} = \begin{bmatrix}
    1 \\
    0
    \end{bmatrix} \otimes \begin{bmatrix}
        0 \\
        1
        \end{bmatrix} = \begin{bmatrix}
            1 \begin{bmatrix}
                0 \\
                1
                \end{bmatrix} \\
            0 \begin{bmatrix}
                0 \\
                1
                \end{bmatrix}
            \end{bmatrix} = \begin{bmatrix}
                0 \\
                1 \\
                0 \\
                0
                \end{bmatrix}
\end{mathpar}

Computation is performed by the application of unitary quantum gates, such as the NOT gate (\textbf{X}), the Hadamard (\textbf{H}) gate, or the controlled-NOT (\textbf{CX}) gate, that can be represented as square matrices. Mathematically, a gate application is equivalent to multiplying the matrix representing the gate with the state vector of the qubit(s). \cref{table:gates} shows some commonly used gates along with their circuit notation and matrix representations. For example:
\begin{mathpar}
X\ket{0} = \begin{bmatrix}
    0 & 1\\
    1 & 0
    \end{bmatrix} \begin{bmatrix}
        1 \\
        0
        \end{bmatrix} = \begin{bmatrix}
            0 \\
            1
            \end{bmatrix}
\end{mathpar}

\begin{table}
\centering
\begin{tabular}{ c c c c }
 \textbf{Gate} & \textbf{Name} & \textbf{Notation} & \textbf{Matrix}\\[2mm]
 I & Identity & \begin{tikzcd} & \gate{I} & \qw \end{tikzcd} & $\begin{bmatrix}
1 & 0 \\
0 & 1
\end{bmatrix}$\\[4mm]
 X & Pauli X & \begin{tikzcd}
     & \gate{X} & \qw
 \end{tikzcd} & $\begin{bmatrix}
0 & 1\\
1 & 0
\end{bmatrix}$\\[4mm]
 Y & Pauli Y & \begin{tikzcd}
     & \gate{Y} & \qw
 \end{tikzcd} & $\begin{bmatrix}
0 & -i\\
i & 0
\end{bmatrix}$\\[4mm]
 Z & Pauli Z & \begin{tikzcd}
     & \gate{Z} & \qw
 \end{tikzcd} & $\begin{bmatrix}
1 & 0 \\
0 & -1
\end{bmatrix}$\\[4mm]
 H & Hadamard & \begin{tikzcd}
     & \gate{H} & \qw
 \end{tikzcd} & $\frac{1}{\sqrt{2}}\begin{bmatrix}
1 & 1 \\
1 & -1
\end{bmatrix}$\\[4mm]
 CX & controlled-NOT & \begin{tikzcd}
     & \ctrl{1} & \qw \\
     & \targ{} & \qw
 \end{tikzcd} & $\begin{bmatrix}
1 & 0 & 0 & 0\\
0 & 1 & 0 & 0\\
0 & 0 & 0 & 1\\
0 & 0 & 1 & 0
\end{bmatrix}$\\[8mm]
 CZ & controlled-Z & \begin{tikzcd}
    & \ctrl{1} & \qw \\
    & \gate{Z} & \qw
 \end{tikzcd} & $\begin{bmatrix}
1 & 0 & 0 & 0\\
0 & 1 & 0 & 0\\
0 & 0 & 1 & 0\\
0 & 0 & 0 & -1
\end{bmatrix}$
\end{tabular}
\caption{Common unitary gates in quantum computation}
\label{table:gates}
\end{table}

This basic unit of information is not the only reason why quantum computing is fundamentally different from classical computing---certain quantum mechanical phenomena are highly counterintuitive! One such phenomenon is \textit{quantum entanglement}, which, at its most basic level, involves a correlation between two qubits such that when one is measured, the outcome obtained necessarily influences the measurement outcome of the other qubit (even if the two qubits are far apart). For example, a simple but well-known fully entangled quantum state over two qubits is the first Bell state, which is often written as $\ket{\beta_{00}} = (\ket{00}+\ket{11})/\sqrt{2}$, where the first $0$ in $\ket{00}$ corresponds to the first qubit and the second $0$ to the second qubit and similarly for $\ket{11}$. In this state, the probability of obtaining either $00$ or $11$ for the pair of qubits on measurement is 1/2 each. That is, if we were to measure only the first qubit and obtain $0$, it is guaranteed that on measurement, we will find the other qubit to also be in the state $0$. When a qubit is not entangled, it is said to be in a \textit{separable} state.

We refer the reader to the standard textbook by \textcite{nielsen2010} for more background. We also recommend the excellent series of essays by \textcite{matuschak2019} for a gentle introduction and to build intuition.

\subsection{Quantum Hoare logic with ghost variables}

We need a predicate logic analogous to Separation Logic~\parencite{reynolds2002} for quantum computation to be able to reason about only the interesting portions of the quantum state while still ensuring the correctness of non-local effects such as entanglement. Various quantum Hoare logics that have been proposed~\parencite{aqhl2019,floydhoare2012,wpe2016} do not support the frame rules that provide Separation Logic its power. The closest approach is that taken by \textcite{unruh2019} that allows predicates on the quantum state similar to what we would like in our setting, such as whether certain qubits are in a classical state, or uniform superposition, entangled or separable.\footnote{Thanks to Brad Lackey for pointing out that Unruh's use of auxiliary systems to express properties of states is equivalent to the notion of \textit{purification}~\parencite[110]{nielsen2010} in quantum information theory.} In this section, we provide an introduction to Unruh's \citetitle{unruh2019} that we use in our predicate language.

Let us make precise some mathematical definitions that we need in this section.

\begin{definition}[Vector Subspace]
	A subspace of a vector space is a nonempty subset of its vectors that is closed under the operations of vector addition and scalar multiplication.
\end{definition}

\begin{remark}
    All vector spaces are equipped with at least two (\textit{trivial}) subspaces: the singleton set, $\{ 0 \}$, consisting of the zero vector and the vector space itself.
\end{remark}

\begin{definition}[Span of state vectors]
	For any subset $P$ in a Hilbert space $\mathcal{H}$, written ($P \subseteq \mathcal{H}$), $\mathit{span}\{P\}$ is the smallest (closed) subspace of $\mathcal{H}$ containing $P$. In other words, $\mathit{span}\{P\}$ is the set of linear combinations of vectors in $P$.
\end{definition}

A core idea behind various quantum logics that comes from \textcite{logicqm36} is that closed subspaces of a state space can be thought of as logical propositions with the inclusion relationship between them forming a partial order. Unruh's quantum Hoare logic defines syntactic sugar for propositions and predicates on subspaces that we will find useful and that we describe next:

\begin{itemize}
	\item $\top$ is the proposition corresponding to the complete state space, $\mathcal{H}$, under consideration. It is logically equivalent to the always true proposition.
	\item $\bot$ is the proposition corresponding to the zero-dimension subspace, $\{0\}$, and is logically equivalent to the absurd proposition.
	\item $P \wedge Q$ is the set intersection of the two subspaces: $P \cap Q$. This is equivalent to the logical \textit{and} connective.
	\item $P \vee Q$ is the sum (linear combination) of the subspaces $P$ and $Q$. In other words, it is the span of the union of the two subspaces: $\mathit{span}\{P \cup Q\}$. This is equivalent to the logical \textit{or} connective.
	\item $U \cdot P$ corresponds to the application of the unitary operator $U$ to a given subspace $P$. It is sugar for the subspace $\{U \psi: \psi \in P\}$.
	\item\label{itm:partnot} $X \in_q P$ lets us specify that the qubits referenced in the list of variables $X$ lie in the subspace P (of a dimension same as or larger than $2^n$, where $n$ is the number of qubits in $X$). In other words, if $V$ is the set of all quantum variables, then this predicate corresponds to the complete quantum state being $P \otimes \mathcal{H}[V \backslash X]$ where the brackets specify the domain of qubits that we are referring to and the backslash is the set difference operator.
	\item $X \equiv_q Y$, called the quantum equality between two disjoint lists of variables $X$ and $Y$ (of the same dimension), states that $X$ and $Y$ are quantumly equal if and only if swapping them in the complete quantum state will not change the state.
	\item $X \equiv_{cl} Y$, called the classical equality between two disjoint lists of variables $X$ and $Y$ (of the same dimension), states that $X$ and $Y$ are classically equal if and only if measuring them in the computational basis will always give the same outcome.
	\item $X =_q \psi$ specifies that variables $X$ are in the specific state $\psi$. This is a sugar for $X \in_q \mathit{span}\{\psi\}$.
\end{itemize}

The last predicate, for example, lets us write $(a, b) =_q (\ket{00}+\ket{11})/\sqrt{2}$ to specify that quantum variables (or qubits) $a, b$ are in the first Bell state. It is easy to see that the span of the given state vector, $\mathit{span}\{(\ket{00}+\ket{11})/\sqrt{2}\}$, only includes the set of all of its scalar multiples. Since state vectors need to obey the normalization constraint on their amplitudes, the scalar multiple can only be a complex number of the form $e^{i\theta}$ whose absolute value is always 1.\footnote{This scalar, called the \textit{global phase factor}, can be ignored in most cases as it does not affect the computation.} Thus, the span of a single vector is just that vector, and hence, we can express that certain quantum variables correspond to a specific state vector using the notion of subspaces as propositions.

Unruh's logic further utilizes ghost variables (also known as logical variables) that can only appear in propositions to make it easy to state properties about the local state when a variable corresponding to that state may not be in scope. For example, it becomes possible to state that a single qubit input to a program is entangled with another qubit, even though we may never see that qubit locally. This is quite useful when proving the quantum teleportation protocol that we will see later.

Ghost variables in Unruh's set-up are identified by their mnemonic names. He uses $g$ \& $G$ for a ghost variable and a list of ghost variables respectively, and similarly $e$ \& $E$ for entangled ghosts and $u$ \& $U$ for unentangled ghosts. For example, if we would like to use the qubit $b$ from the above qubit pair individually, we can specify that it is in the first Bell state as $(e, b) =_q (\ket{00}+\ket{11})/\sqrt{2}$, where $e$ is a fresh entangled ghost variable. Intuitively, a ghost variable can assume whatever state is needed to make a given proposition true.

Unruh further defines some syntactic sugar for predicates that include ghosts:

\begin{itemize}
    \item $\mathbf{uniform}(X)$, which states that $X$ is uniformly distributed. It is sugar for $(X, e) =_q \sum_i \frac{1}{\sqrt{m}} (\ket{i} \otimes \ket{i})$ where $e$ is a fresh entangled ghost and $i$ ranges from bitvectors $0$ to $m-1$.
    Note that it is not possible to express this without using ghosts. Consider, for example, a single qubit in $\mathbb{C}^2$ in a uniform distribution of its basis states. The only subspace that includes both $\ket{0}$ and $\ket{1}$ is $\top$.
	\item $\mathbf{separable}(X)$, which states that $X$ is separable, that is, not entangled with any other system. This is sugar for $X \equiv_q u$ where is $u$ is a fresh unentangled ghost.
	\item $\mathbf{class}(X)$, which states that $X$ holds a classical state. This is sugar for $X \equiv_{cl} u$ where $u$ is a fresh unentangled ghost.
\end{itemize}

Unruh demonstrates the usefulness of these predicates over a simple imperative quantum language that includes primitives for initialization, unitary application, and measurement. We take inspiration from his logic and inference rules while designing the specification language of QHTT that we describe in the next section.

For reference, his rules for the unitary application and initialization commands are reproduced here:
\begin{mathpar}
    \inferrule[Unitary]
    {}
    {\HoareT{P}{\textbf{apply}\ U\ \textbf{to}\ X}{(U\ \textbf{on}\ X) \cdot P}} \and
    \inferrule[Initialization]
    {}
    {\HoareT{P}{\textbf{init}\ x}{P[x \rightarrow e],\ x =_q \ket{0}}}
\end{mathpar}

The unitary application rule simply produces the strongest postcondition from the precondition $P$ by applying the given operator to the chosen qubit(s), $X$, in the subspace $P$. In $U\ \textbf{on}\ X$, $U$ is appropriately padded with the identity operation so that the operation $U \otimes I_{[V \backslash X]}$ is applied to the quantum state corresponding to all quantum variables, $V$. The initialization rule substitutes a fresh entangled ghost $e$ for the existing quantum variable $x$ and then initializes $x$ to $\ket{0}$. As we will see, our unitary application rule is quite similar, but our initialization command has different semantics and hence a different rule. Another distinction from Unruh's work is that he considers an infinite-dimensional Hilbert space, while in QHTT we consider a finite number of qubits that are determined via the program syntax.

\section{Quantum Hoare Type Theory}
To design a Hoare Type Theory for quantum computation, we need to consider some assumptions about the quantum state and its programming model.

Each program is assumed to have access to a pool of countably infinite number of logical qubits that can be initialized on demand and are returned to the pool (discarded) on measurement. With ``logical'', we mean qubits with no errors; such qubits may be realized using quantum error correction techniques. We define primitive quantum commands for qubit initialization, unitary application, and measurement next.

\subsection{Primitive quantum commands}
\label{sec:primcmds}

Here we show our primitive quantum commands that are inspired by similar primitives in \citetitle{qio}~\parencite{qio} along with the Hoare types that encode the specifications for these commands. The predicates that we see are the same as we saw in the previous section. Each predicate implicitly takes the current local state of the program as an argument. These specifications are given in a small footprint style.

\subsubsection{Initialization}
\begin{lstlisting}[language=QHaskell]
init : (b: bit) -> QST (q: qbit)
                       (requires {⊤})
                       (ensures  {q =q |b⟩})
\end{lstlisting}

We can initialize one qubit at a time in either of the two computational basis states, $\ket{0}$ or $\ket{1}$. For example, the initialization command \lstinline[language=QHaskell]{x <- init 0} allocates a fresh qubit from the global pool, initializes it to $\ket{0}$ and returns a reference to the qubit and binds it to $x$.

The precondition is true for any local state, which suggests that this command may be called in any context. The postcondition states that the modified local state includes a new binding for the newly allocated qubit.

Note that our initialization command is different from Unruh's work discussed in the previous section. The argument to our command is the initial classical value, while in Unruh's $init$ the argument is a qubit that gets initialized to $\ket{0}$.

\subsubsection{Measurement}

\begin{lstlisting}[language=QHaskell]
meas : (q: qbit) -> QST (x: bit)
                        {ψ: vector, e_x: qbit}
                        (requires {(q⊗ℋ[V#q]) =q ψ})
                        (ensures  {class(q) ∧ (e_x⊗ℋ[V#q]) =q ψ})
\end{lstlisting}

We assume all measurements to be in the computational basis. The measurement command, in effect, returns the qubit to the pool after returning the result of the measurement as a bit. We record the distribution of the outcome by introducing an entangled ghost variable that replaces the measured qubit in the global state. How this helps will be clear when we work through some examples later in the document. We continue to use the state space partition notation ($q\otimes\mathcal{H}[V\backslash q]$) first used in \cref{itm:partnot} to refer to only the local state under consideration. Note that this notation should not be confused to mean that $q$ is separable from the rest of the system (it may or may not be).

\subsubsection{Unitary application}

\begin{lstlisting}[language=QHaskell]
apply _ to _ : (g: unitary) ->
               qs: (qbit⊗qbit) ->
               QST (_: unit)
                   {P: prop}
                   (requires {P})
                   (ensures  {(g on qs) . P})
\end{lstlisting}

For convenience, we only show the rule for two-qubit gate application, but we support both one- and two-qubit gates. Since it modifies the state in place, the unitary application command returns the trivial unit output. The pre- and postconditions are the same as Unruh's rule for unitary application command.

These primitive quantum commands and a specification language based on the predicates proposed by \textcite{unruh2019} make the core of what is different in our Quantum Hoare Type Theory from the original formulation of HTT.

\subsection{Syntax}

Now, we are in a good position to talk about the raw syntax of QHTT. The following grammar shows the available base types and type constructors in the language:

\nonterms{Typ}

The base type $bit$ inhabits the values 0 and 1, but the type $qbit$ is just a reference to a logical qubit. The state is represented using the type $vector$ that may further need the type $complex$ to represent arbitrary quantum states. These last two types are only included for use in the specification language and are not to be used by the programming syntax. The type $unitary$ includes all basic one- and two-qubit gates (such as the Pauli gates and CX) that we include in the language. The propositions formed using the specification language are given the $prop$ type. $\otimes$ is the pair type constructor. Note that we use the F*-inspired syntax \lstinline[language=QHaskell]!QST (x:A) {Γ} (requires {P}) (ensures {Q})! in examples, which is just syntactic sugar for the Quantum Hoare Type shown in the formal syntax above. The optional context, $\Gamma$, is used to bind ghost variables and propositions that may appear in the pre- and postcondition to relate the beginning and the end state, as we saw in the rules for measurement and unitary application.

Next, we show the grammar for the terms of the language that include values for the base types and introduction and elimination forms for the type constructors:

\nonterms{Tm}

We note that we do not provide a concrete syntax for specifying quantum state vectors that are only used in the propositions. The term $do\ E$ represents a suspended computation and is the introduction form for the Hoare type.

We have already seen the primitive commands:

\nonterms{Cmd}

We did not mention the classical control construct if-then-else earlier. It branches to $E_1$ or $E_2$ based on the truth value of the bit $M$.

We next show the predicates of the specification language from \textcite{unruh2019}.

\nonterms{Prop}

\begin{table}
    \centering
    \begin{tabular}{ c c c }
        \textbf{Connective} & \textbf{Formal name} & \textbf{Informal name} \\
        $\bot$ & contradiction & false \\
        $\top$ & true & true\\
        $\wedge$ & conjunction & and\\
        $\vee$ & disjunction & or\\
        $\Rightarrow$ & implication & implies\\
    \end{tabular}
    \caption{Propositional logic connectives}
    \label{table:proplogic}
\end{table}

Apart from the quantum predicates that we see above, we also have access to the usual connectives from propositional logic, which we summarize in \cref{table:proplogic}. Some of these connectives have the same symbols as the quantum predicates, but their meaning will always be clear from the context. Additionally, we use $x =_c v$ to denote that a classical bit $x$ is equal to a value $v$.

Throughout the type-checking process, we need to maintain some environment information in the form of contexts that track variable bindings and propositions available at a point in the program:

\nonterms{Ctx}

These contexts are especially relevant when calling a quantum function as the function may require the binding of the ghost variables in its specification from the environment that it is called in. Similarly, a \textit{merge} of environments may be needed when a function call returns to determine the most precise propositions at the place of return.

Finally, we show the computations that represent the monadic unit and three kinds of bind operations:

\nonterms{Comp}

The complete grammar and typing rules are included in \cref{app:grammar,app:rules} of this document for easy reference. Most of the typing rules are similar to the classical HTT.

\subsection{Examples}

In this section, we show some simple examples along with their specifications as types. We also show the verification steps at each point in the program.

\subsubsection{Bell states}

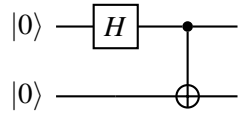
\begin{figure}
    \centering
    \begin{tikzpicture}
        \node[scale=1.0] {
            \begin{quantikz}
                \lstick{$\ket{0}$} & \gate{H} & \ctrl{1} & \qw \\
                \lstick{$\ket{0}$} & \qw      & \targ{}  & \qw \\
            \end{quantikz}
        };
    \end{tikzpicture}
    \caption{Circuit to prepare the first Bell state}
    \label{fig:bell00}
\end{figure}

Our simplest example involves the creation of a Bell (or EPR) state, which is one of the four maximally entangled quantum states of two qubits. Specifically, we create a circuit (\cref{fig:bell00}) to produce the first Bell state, which we write in the mnemonic notation as $\ket{\beta_{00}} = (\ket{00}+\ket{11})/\sqrt{2}$.

\begin{lstlisting}[language=QHaskell,float,label=lst:bell,lastline=9,caption=Generating the first Bell state]
bell00 : unit -> QST (a, b): (qbit⊗qbit)
                     (requires {⊤})
                     (ensures  {(a,b) =q |β00⟩})
bell00 = \x.do
            a <- init 0
            apply H to a
            b <- init 0
            apply CX to (a, b)
            return (a, b)
\end{lstlisting}

\begin{lstlisting}[language=QHaskell,float,label=lst:bella,firstline=11,caption=Bell state program annotated with propositions]
bell00 : unit -> QST (a, b): (qbit⊗qbit)
                     (requires {⊤})
                     (ensures  {(a,b) =q |β00⟩})
bell00 = \x.do
            -- ⊤
            a <- init 0
            -- a =q |0⟩
            apply H to a
            -- (H on a) . (a =q |0⟩)
            -- ⇒ a =q |+⟩
            b <- init 0
            -- a =q |+⟩ ∧ b =q |0⟩
            apply CX to (a, b)
            -- (CX on (a, b)).(a =q |+⟩ ∧ b =q |0⟩)
            -- ⇒ (a,b) =q |β00⟩
            return (a, b)
\end{lstlisting}

\cref{lst:bell} shows a circuit that corresponds to a suspended quantum computation that can be composed with other circuits. It takes no input and returns two qubits as output. The specification of the program is in its type (the first three lines). Here we see our first two propositions: top, $\top$, which is satisfied by any state space, and the $X =_q \psi$ predicate that is satisfied by quantum variables, $X$, that lie in the span of the given state $\psi$. Intuitively, this circuit does not require any inputs and can work in any given state space, and produces two qubits that are maximally entangled in the first Bell state $\ket{\beta_{00}}$.

\cref{lst:bella} shows how we can prove whether this program conforms to the given specification. The initialization command (at lines 6 and 11) allocates a new qubit, initializes it to $\ket{0}$, and binds it to the variable on the left side of $\leftarrow$. Each occurrence of initialization adds its strongest postcondition, as shown in the form of comments. At line 8, the Hadamard gate application on $\ket{0}$ results in the equal superposition state, $\ket{+} = \frac{1}{\sqrt{2}}(\ket{0}+\ket{1})$. The strongest postcondition for a unitary application simply involves applying the operator to the subspace specified by the precondition (on lines 9 and 14). Note that we use the $\Leftrightarrow$ symbol in these annotations for a simplification step. Finally, the return operation lifts the return value to the Quantum Hoare type along with its strongest postcondition (which is trivially the same as the precondition).

\subsubsection{Quantum teleportation}
\label{sec:teleport}

\begin{figure}
    \centering
    \begin{tikzpicture}
        \node[scale=1.0] {
            \begin{quantikz}
                \lstick{$\psi$}                & \ctrl{1} & \gate{H} & \meter{} & \cw        & \cwbend{2} \\
                \lstick[wires=2]{$\ket{\beta_{00}}$} & \targ{}  & \qw      & \meter{} & \cwbend{1} \\
						       \qw      & \qw      & \qw      & \qw        & \gate{X}   & \gate{Z} & \qw \rstick{$\psi$}\\
            \end{quantikz}
        };
    \end{tikzpicture}
    \caption{Quantum teleportation circuit}
    \label{fig:teleport}
\end{figure}
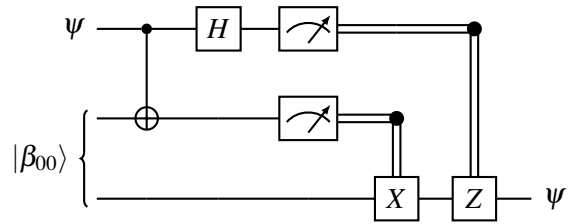

\begin{lstlisting}[language=QHaskell,float,label=lst:tele,lastline=13,caption=Quantum teleportation]
teleport : q: qbit -> QST b: qbit
                          {ψ: vector}
                          (requires {q =q ψ})
                          (ensures  {class(q) ∧ b =q ψ})
teleport = \q.do
              (a, b) <- bell00 ()
              apply CX to (q, a)
              apply H to q
              x <- meas q
              y <- meas a
              apply CX to (y, b)
              apply CZ to (x, b)
              return b
\end{lstlisting}

Our second example shows how we can compose circuits defined in QHTT together and still maintain correctness. \Cref{fig:teleport} shows the circuit for quantum teleportation~\parencite{bennett93} and \cref{lst:tele} shows the corresponding code, which uses the Bell pair generation circuit we defined above. This example also demonstrates how we can bind ghost variables in the Hoare type: $\{\psi: \mathit{vector}\}$ at line 2 states that $\psi$ is a ghost variable of \textit{vector} type. Recall that ghost variables can only be used in logical propositions and are not part of the program. We place the context declaring the ghost variables right after the output type to signal that the ghost variables are only bound in the \textit{requires} and \textit{ensures} constructors that follow next in the Hoare type. We see another predicate $\mathbf{class}(q)$ here that states that the given qubit $q$ is in a classical state, which is what we expect during the quantum teleportation protocol. Or there would be no way to tell from the specification whether the implementation just returned the input qubit as the output.

\begin{lstlisting}[language=QHaskell,float,label=lst:telea,firstline=15,caption=Annotated quantum teleportation program]
teleport : q: qbit -> QST b: qbit
                          {ψ: vector}
                          (requires {q =q ψ})
                          (ensures  {class(q) ∧ b =q ψ})
teleport = \q.do
              -- q =q ψ
              -- ⇒ q =q α|0⟩+β|1⟩
              (a, b) <- bell00 ()
              -- q =q ψ ∧ (a, b) =q |β00⟩
              -- ⇒ (q, a, b) =q 1/√2(α|0⟩(|00⟩+|11⟩)+β|1⟩(|00⟩+|11⟩))
              apply CX to (q, a)
              -- (CX on (q, a)) . (q =q ψ ∧ (a, b) =q |β00⟩)
              -- ⇒ (q, a, b) =q 1/√2(α|0⟩(|00⟩+|11⟩)+β|1⟩(|10⟩+|01⟩))
              apply H to q
              -- (H on q) . ((q, a, b) =q 1/√2(α|0⟩(|00⟩+|11⟩)+β|1⟩(|10⟩+|01⟩)))
              -- ⇒ (q, a, b) =q 1/2(α(|0⟩+|1⟩)(|00⟩+|11⟩) + β(|0⟩-|1⟩)(|10⟩+|01⟩))
              -- ⇒ (q, a, b) =q 1/2(|00⟩(α|0⟩+β|1⟩) + |01⟩(α|1⟩+β|0⟩)
              --                   + |10⟩(α|0⟩-β|1⟩) + |11⟩(α|1⟩-β|0⟩))
              x <- meas q
              -- class(q) ∧
              -- (e_x, a, b) =q 1/2(|00⟩(α|0⟩+β|1⟩) + |01⟩(α|1⟩+β|0⟩)
              --                 + |10⟩(α|0⟩-β|1⟩) + |11⟩(α|1⟩-β|0⟩))
              y <- meas a
              -- class(q) ∧ class(a) ∧
              -- (e_x, e_y, b) =q 1/2(|00⟩(α|0⟩+β|1⟩) + |01⟩(α|1⟩+β|0⟩)
              --                  + |10⟩(α|0⟩-β|1⟩) + |11⟩(α|1⟩-β|0⟩))
              apply CX to (y, b)
              -- class(q) ∧ class(a) ∧
              -- (e_x, e_y, b) =q 1/2(|00⟩(α|0⟩+β|1⟩) + |01⟩(α|0⟩+β|1⟩)
              --                  + |10⟩(α|0⟩-β|1⟩) + |11⟩(α|0⟩-β|1⟩))
              apply CZ to (x, b)
              -- class(q) ∧ class(a) ∧
              -- (e_x, e_y, b) =q 1/2(|00⟩(α|0⟩+β|1⟩) + |01⟩(α|0⟩+β|1⟩)
              --                  + |10⟩(α|0⟩+β|1⟩) + |11⟩(α|0⟩+β|1⟩))
              -- ⇒ (e_x, e_y) =q 1/2(|00⟩+|01⟩+|10⟩+|11⟩) ∧ b =q α|0⟩+β|1⟩
              return b
              -- class(q) ∧ b =q α|0⟩+β|1⟩
              -- ⇒ class(q) ∧ b =q ψ
\end{lstlisting}

Verification steps for this program are shown in \cref{lst:telea}. In line 8, we see the call to the \texttt{bell00} function. Since the precondition of the function is trivial, there is nothing unusual, but after the function call, the calling environment is merged with the postcondition of the function. We see the expected postcondition on the newly bound qubit pair, $(a, b)$, just before the simplification step. In line 19, something more interesting happens! When we measure a quantum variable (in the default computational basis), its state is collapsed to either the classical bit 0 or 1 based on the distribution of either outcome. In our logic, however, both the outcome and the distribution are still maintained. This is achieved by replacing the measured qubit (which is implicitly discarded in the measurement operation) in the state space under consideration with an entangled ghost variable. Furthermore, we maintain the correspondence between the classical bit that stores the measurement result and the ghost variable that replaced its corresponding qubit. So at line 21, $e_x$ corresponds to the classical bit $x$. Similarly, in lines 24--26, $e_y$ is introduced corresponding to $y$. This helps us in the next two steps of the program where operations controlled on the previous measurement outcomes are applied. We recover the original state $\psi$ at the end. Note that we use the comma symbol interchangeably with the conjunction symbol in the propositions.

This version of teleportation may not seem very interesting because it is essentially equivalent to teleportation performed without measurements (using the principle of deferred measurements). Hence, we turn to a more realistic variant in our next example.

\subsubsection{Modular teleportation with classical control}

As the teleportation protocol is usually stated, we have two agents Alice and Bob, who share an entangled Bell pair at the beginning of the protocol. Alice wants to communicate a message encoded in a qubit to Bob. Alice performs a Bell measurement on her two qubits, the one that encodes the message and the other which is part of the entangled Bell pair. Then she sends the result of this measurement---the two bits---to Bob. Bob receives those bits and performs correction steps on his qubit to obtain the message. We can naturally split the teleportation circuit into three parts: Bell state preparation, Alice's portion of the circuit, and Bob's portion of the circuit.

\begin{lstlisting}[language=QHaskell,float,label=lst:alice,caption=Alice's portion of teleportation,lastline=14]
alice : (q, a): (qbit⊗qbit) ->
        QST (x, y): (bit⊗bit)
            {e: qbit, e_x: qbit, e_y: qbit, α: complex, β: complex}
            (requires {q =q α|0⟩+β|1⟩ ∧ (a, e) = |β00⟩})
            (ensures  {class(q) ∧ class(a) ∧
                       (e_x, e_y, e) =q 1/2(|00⟩(α|0⟩+β|1⟩) + |01⟩(α|1⟩+β|0⟩)
                                        + |10⟩(α|0⟩-β|1⟩) + |11⟩(α|1⟩-β|0⟩))})
alice = \qa.do
            (q, a) = qa
            apply CX to (q, a)
            apply H to q
            x <- meas q
            y <- meas a
            return (x, y)
\end{lstlisting}

\begin{lstlisting}[language=QHaskell,float,label=lst:bob,caption=Bob's portion of teleportation,firstline=45,lastline=55]
bob : (x, y, b): (bit⊗bit⊗qbit) ->
      QST q: qbit
          {c: complex, d: complex}
          (requires {b =q c|0⟩+d|1⟩})
          (ensures  {y =c 0 => q =q (c|0⟩ + (-1)^x d|1⟩) ∧
                     y =c 1 => q =q (c|1⟩ + (-1)^x d|0⟩)})
bob = \xyb.do
          (x, y, b) = xyb
          if y then apply X to b else ()
          if x then apply Z to b else ()
          return b
\end{lstlisting}

\begin{lstlisting}[language=QHaskell,float,label=lst:mtele,caption=Modular teleportation,firstline=74,lastline=82]
teleport : q: qbit -> QST b: qbit
                          {ψ: vector}
                          (requires {q =q ψ})
                          (ensures  {class(q) ∧ b =q ψ})
teleport = \q.do
              (a, b) <- bell00 ()
              (x, y) <- alice (q, a)
              b <- bob (x, y, b)
              return b
\end{lstlisting}

\cref{lst:alice,lst:bob,lst:mtele} show the implementation of the latter two of these parts (we have already seen Bell state preparation) and the final \texttt{teleport} circuit (that composes the three parts together) in the form of separate modules (functions) along with their specifications as types. Alice's circuit requires that her second input qubit be entangled in the Bell state. This precondition shows our first example of using Unruh's entangled ghost variable, $e$, in our specification, which is declared in the context right before the \textit{requires} construct along with other ghost variables. It further ensures that her two input qubits are both consumed in the process as they are asserted to be classical in the postcondition. Since Alice is dealing with entanglement, it is important to maintain the distribution over all qubits involved, including the ghost qubit $e$ and the newly created ghosts $e_x$ and $e_y$ because of her measurement operations.

Bob's specification shows what correction is applied if his qubit is in an arbitrary state. Note that \texttt{bob} uses the classical control construct if-then-else in this variant of the circuit. Moreover, note that to distinguish between the input and output of the qubit, we are using two different names, but both $b$ and $q$ refer to the same qubit. The specification for the \texttt{teleport} function is the same as before, but the implementation is now much shorter.

\begin{lstlisting}[language=QHaskell,float,label=lst:alicea,caption=Annotated code for alice,firstline=16,lastline=43]
alice : (q, a): (qbit⊗qbit) ->
        QST (x, y): (bit⊗bit)
            {e: qbit, e_x: qbit, e_y: qbit, α: complex, β: complex}
            (requires {q =q α|0⟩+β|1⟩ ∧ (a, e) = |β00⟩})
            (ensures  {class(q) ∧ class(a) ∧
                       (e_x, e_y, e) =q 1/2(|00⟩(α|0⟩+β|1⟩) + |01⟩(α|1⟩+β|0⟩)
                                        + |10⟩(α|0⟩-β|1⟩) + |11⟩(α|1⟩-β|0⟩))})
alice = \qa.do
            (q, a) = qa
            -- q =q α|0⟩+β|1⟩ ∧ (a, e) =q |β00⟩}
            -- ⇒ (q, a, e) =q 1/√2(α|0⟩(|00⟩+|11⟩)+β|1⟩(|00⟩+|11⟩))
            apply CX to (q, a)
            -- (CX on (q, a)) . (q =q ψ ∧ (a, e) =q |β00⟩)
            -- ⇒ (q, a, e) =q 1/√2(α|0⟩(|00⟩+|11⟩)+β|1⟩(|10⟩+|01⟩))
            apply H to q
            -- (H on q) . ((q, a, e) =q 1/√2(α|0⟩(|00⟩+|11⟩)+β|1⟩(|10⟩+|01⟩)))
            -- ⇒ (q, a, e) =q 1/2(α(|0⟩+|1⟩)(|00⟩+|11⟩) + β(|0⟩-|1⟩)(|10⟩+|01⟩))
            -- ⇒ (q, a, e) =q 1/2(|00⟩(α|0⟩+β|1⟩) + |01⟩(α|1⟩+β|0⟩)
            --                   + |10⟩(α|0⟩-β|1⟩) + |11⟩(α|1⟩-β|0⟩))
            x <- meas q
            -- class(q) ∧
            -- (e_x, a, e) =q 1/2(|00⟩(α|0⟩+β|1⟩) + |01⟩(α|1⟩+β|0⟩)
            --                 + |10⟩(α|0⟩-β|1⟩) + |11⟩(α|1⟩-β|0⟩))
            y <- meas a
            -- class(q) ∧ class(a) ∧
            -- (e_x, e_y, e) =q 1/2(|00⟩(α|0⟩+β|1⟩) + |01⟩(α|1⟩+β|0⟩)
            --                  + |10⟩(α|0⟩-β|1⟩) + |11⟩(α|1⟩-β|0⟩))
            return (x, y)
\end{lstlisting}

\begin{lstlisting}[language=QHaskell,float,label=lst:boba,caption=Annotated code for bob,firstline=57,lastline=72]
bob : (x, y, b): (bit⊗bit⊗qbit) ->
      QST q: qbit
          {c: complex, d: complex}
          (requires {b =q c|0⟩+d|1⟩})
          (ensures  {y =c 0 => q =q (c|0⟩ + (-1)^x d|1⟩) ∧
                     y =c 1 => q =q (d|0⟩ + (-1)^x c|1⟩)})
bob = \xyb.do
          (x, y, b) = xyb
          -- b =q c|0⟩+d|1⟩
          if y then apply X to b else ()
          -- y =c 0 => b =q (c|0⟩ + d|1⟩) ∧
          -- y =c 1 => b =q (c|1⟩ + d|0⟩)
          if x then apply Z to b else ()
          -- y =c 0 => b =q (c|0⟩ + (-1)^x d|1⟩) ∧
          -- y =c 1 => b =q (c|1⟩ + (-1)^x d|0⟩)
          return b
\end{lstlisting}

\begin{lstlisting}[language=QHaskell,float,label=lst:mtelea,caption=Annotated code for modular teleportation,firstline=84]
teleport : q: qbit -> QST b: qbit
                          {ψ: vector}
                          (requires {q =q ψ})
                          (ensures  {class(q) ∧ b =q ψ})
teleport = \q.do
              -- q =q ψ
              -- ⇒ q =q α|0⟩+β|1⟩
              (a, b) <- bell00 ()
              -- q =q ψ ∧ (a, b) =q |β00⟩
              -- ⇒ (q, a, b) =q 1/√2(α|0⟩(|00⟩+|11⟩)+β|1⟩(|00⟩+|11⟩))
              (x, y) <- alice (q, a)
              -- class(q) ∧ class(a) ∧
              -- (e_x, e_y, b) =q 1/2(|00⟩(α|0⟩+β|1⟩) + |01⟩(α|1⟩+β|0⟩)
              --                  + |10⟩(α|0⟩-β|1⟩) + |11⟩(α|1⟩-β|0⟩))
              b <- bob (x, y, b)
              -- Merge:
              -- class(q) ∧ class(a) ∧
              -- (e_x, e_y, b) =q 1/2(|00⟩(α|0⟩+β|1⟩) + |01⟩(α|1⟩+β|0⟩)
              --                  + |10⟩(α|0⟩-β|1⟩) + |11⟩(α|1⟩-β|0⟩))
              -- and
              -- y =c 0 => b =q (c|0⟩ + (-1)^x d|1⟩) ∧
              -- y =c 1 => b =q (c|1⟩ + (-1)^x d|0⟩)
              -- ⇒
              -- class(q) ∧ class(a) ∧
              -- (e_x, e_y, b) =q 1/2(|00⟩(α|0⟩+β|1⟩) + |01⟩(α|0⟩+β|1⟩)
              --                  + |10⟩(α|0⟩+β|1⟩) + |11⟩(α|0⟩+β|1⟩))
              -- ⇒ (e_x, e_y) =q 1/2(|00⟩+|01⟩+|10⟩+|11⟩) ∧ b =q α|0⟩+β|1⟩
              return b
              -- b =q α|0⟩+β|1⟩ ⇒ b =q ψ
\end{lstlisting}

We show the annotated verification steps in \cref{lst:alicea,lst:boba,lst:mtelea}. The verification conditions generated for \texttt{alice} are almost the same as those generated during the first part of our first teleportation example, except we do not have the qubit $b$ in scope and instead are dealing with its ghost $e$.

In Bob's case, we start with his qubit $b$ in some arbitrary quantum state and perform corrections as specified by the classical conditionals. We do not need entangled ghosts here as Bob is only dealing with two classical bits and his own qubit.

Finally, in the teleport code in \cref{lst:mtelea}, while calling the function \texttt{alice} at line 11, the entangled ghost, $e$, in her type is bound to the corresponding qubit $b$ in the calling environment as they are both part of the Bell pair along with $a$. After substituting $b$ for $e$, we obtain the same propositions as before on return from the function. A non-trivial environment merge is needed after line 15 on return from the function \texttt{bob}. We have annotated this merge operation in more detail in lines 16--22 for clarity. After relating the existing ghosts $e_x$ and $e_y$ in the return environment with the classical bits $x$ and $y$ in the postcondition for Bob, we obtain the correct propositions. Simplification after the merge operation lets us recover the original message at line 27.

\subsubsection{Deutsch's algorithm}
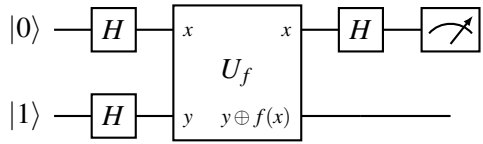
\begin{figure}
    \centering
    \begin{tikzpicture}
        \node[scale=1.0] {
            \begin{quantikz}
                \lstick{$\ket{0}$} & \gate{H} & \gate[wires=2][1.7cm]{U_f}\gateinput{$x$}
                \gateoutput{$x$} & \gate{H} & \meter{} \\
                \lstick{$\ket{1}$} & \gate{H} & \gateinput{$y$}\gateoutput{$y\oplus f(x)$} & \qw & \qw \\
            \end{quantikz}
        };
    \end{tikzpicture}
    \caption{Quantum circuit for Deutsch's algorithm}
    \label{fig:deutsch}
\end{figure}

Our final example in \cref{lst:deutsch} demonstrates that our system can work in a higher-order setting where we can pass a circuit as input to another circuit and reason about the resulting computation. Here $U_f$ is a quantum oracle whose implementation is unknown, but its type specifies everything we know about it---it is parameterized over a classical function, $f$, that takes a bit and returns a bit; it takes two qubits as input and returns two qubits as output; and finally, if the input qubits store classical bits $x$ and $y$, then the output qubits store $x$ and $y \oplus f(x)$. The circuit $U_f$ is used as a black box in the implementation of \citeauthor{deutsch85}'s~\parencite*{deutsch85} algorithm as shown in \cref{fig:deutsch}. The idea is to be able to reason about the algorithm's correctness using the type of the black box alone.

Note that the second argument to the \texttt{deutsch} function carries its Hoare type (as specified above) along, even though we do not show it in the code at line 9 to avoid extra noise in the presentation. At line 25, the first argument, $f$, is passed to the oracle as a parameter without further calls to $f$. At lines 27--28, we show the standard case analysis for the application of the oracle to the given quantum states of $a$ and $b$ depending on the constant or balanced property of $f$. We note that this example requires some manual intervention in interpreting the conclusion of the algorithm as we are not allowed to inspect the behavior of $f$ directly.

\begin{lstlisting}[language=QHaskell,float,label=lst:deutsch,caption=Deutsch's algorithm]
U_f : (f: bit -> bit) ->
      (a, b) : (qbit⊗qbit) ->
      QST (c, d) : (qbit⊗qbit)
          {x: bit, y: bit}
          (requires {(a, b) =q |x,y⟩})
          (ensures  {(c, d) =q |x,y⊕f(x)⟩})

deutsch : (f: bit -> bit) ->
          ((bit -> bit) -> (qbit⊗qbit) -> QST (qbit⊗qbit)) ->
          QST r: bit
              (requires {⊤})
              (ensures  {(r =c 0 => f(0) == f(1)) ∧
                         (r =c 1 => f(0) != f(1)) })
deutsch = \f.\U_f.do
            -- ⊤
            a <- init 0
            -- a =q |0⟩
            b <- init 1
            -- a =q |0⟩ ∧ b =q |1⟩
            apply H to a
            -- (H on q) . (a =q |0⟩ ∧ b =q |1⟩)
            -- ⇒ a =q |+⟩ ∧ b =q |1⟩
            apply H to b
            -- a =q |+⟩ ∧ b =q |-⟩
            (a, b) <- U_f f (a, b)
            -- (U_f on (a, b)) . (a =q |+⟩ ∧ b =q |-⟩)
            -- ⇒ (a =q |+⟩ ∧ b =q |-⟩) -- f(0) == f(1)
            --  ∨ (a =q |-⟩ ∧ b =q |-⟩) -- f(0) != f(1)
            apply H to a
            --   (a =q |0⟩ ∧ b =q |-⟩) -- f(0) == f(1)
            -- ∨ (a =q |1⟩ ∧ b =q |-⟩) -- f(0) != f(1)
            r <- meas a
            -- class(a) ∧
            --   (r =c 0 ∧ b =q |-⟩) -- f(0) == f(1)
            -- ∨ (r =c 1 ∧ b =q |-⟩) -- f(0) != f(1)
            return r
\end{lstlisting}

\section{Discussion}

In \cref{sec:teleport}, when we presented the code for quantum teleportation, it was apparent from its type (specification) that the quantum variables $q$ (input) and $b$ (output) are different because of the proposition $\mathbf{class}(q)$, which suggests that $q$ is discarded in the postcondition. It would be nicer if we could also specify information about the resource usage of the program; in teleportation, for example, an EPR pair gets consumed along with the input qubit.

A limitation of our system is that we currently do not support propositions about the distribution of measurement outcomes. For example, \cref{lst:toss} shows a program (along with its verification steps), which produces a uniformly random bit. At line 10, we can infer a weaker proposition on $q$, $\textbf{uniform}(q)$, knowing its exact state. But the proposition $\textbf{uniform}(b)$ in the postcondition of the program would be invalid as it is only defined for quantum variables and not classical bits.

\begin{lstlisting}[language=QHaskell,float,label=lst:toss,firstline=10,caption=Fair coin toss with an \textcolor{red}{invalid} postcondition]
cointoss : unit -> QST b: bit
                       (requires {⊤})
                       (ensures  {uniform(b)})
cointoss = \x.do
              -- ⊤
              q <- init 0
              -- q =q |0⟩
              apply H to q
              -- q =q 1/√2(|0⟩+|1⟩) ⇒ q =q |+⟩
              -- => uniform(q)
              x <- meas q
              -- class(q) ∧ e_x =q |+⟩
              return x
\end{lstlisting}

We would like to mention that the type system presented in this document is not bidirectional~\parencite{dunfield2019bidirectional} as we forego distinguishing between the introduction and elimination forms of terms while defining our syntax for simplicity. The original presentations of HTT~\parencite{nanevski2008,abspr07} (as do many other type theories) keep that distinction to be able to prove the decidability of type checking and provide a syntax-directed type system to aid implementation. We leave this for future work.

Finally, a note about scalability: there are three important aspects to scalability considerations: the expressive power of the programming language to support writing large quantum programs; compositional reasoning about programs for modularity and reuse of verified code; and lastly but importantly, the manual and computational effort required for verification. We have taken the first and second of these into account by intentionally designing the syntax of our language to be close to Quipper~\parencite{quipper2013}, which is a proven, scalable quantum programming language. In particular, our Deutsch's algorithm example shows that the language supports higher-order programming. We further benefit from building upon HTT that lets us seamlessly compose Hoare types together, hence permitting modular verification as we show in our quantum teleportation example. The third aspect of scalability dealing with the proof burden requires some comparison with existing work.

The full-fledged \citetitle{floydhoare2012}~\parencite{floydhoare2012} or QHL, which builds upon previous work on modeling quantum programs as superoperators~\parencite{Selinger2004} and quantum predicates as Hermitian operators~\parencite{dhondt2006}, suffers from two problems. First, it requires verbose specifications over the complete quantum state because it does not support Separation logic-style frame rules. We avoid this problem by adopting Unruh's variant of QHL that effectively lets us get away with only specifying properties of the local state. Second, the verification process is manual, non-trivial, and does not scale to complicated algorithms such as Shor's~\parencite*{shor94}, as a recent implementation effort using an interactive theorem prover shows~\parencite{liu2019}. A fundamental reason behind this second problem is the representation of quantum propositions as Hermitian operators. Indeed, because of this scalability limitation, there is a recent trend in research based on QHL to move away from general Hermitian operators and towards a special class of propositions based on projection operators~\parencite{aqhl2019,Li2020}, which are equivalent to our notion of closed subspaces as propositions. This prior research and our intention to use a hybrid system like F* (letting us rely more on automation) for implementing QHTT suggest that we are in the right direction. A full account of scalability considerations of our approach requires further investigation, and along with an implementation, we leave it for future work.

\section{Conclusions and perspectives}
\label{sec:conclusion}
In this paper, we describe our work on Quantum Hoare Type Theory, which is a dependently typed functional programming language with support for quantum computation. We use the quantum Hoare logic by \textcite{unruh2019} to compactly specify properties about the quantum state but do it in a type theoretical setting. We further demonstrate the usefulness of our approach using simple examples, including bell state preparation and quantum teleportation.

With future work on code extraction, our approach can become a unified system for programming, specification, and reasoning about quantum computation. There are several other avenues to explore:

\paragraph{Quantum Language} Our programming language currently does not support any iterative or recursive constructs. We would like to add those in the future, along with the reasoning rules needed for loop invariants. Furthermore, currently, it is not possible to construct pure unitary operators directly in our language apart from those provided as constants. As \textcite{qio} show, unitary operations form an algebraic monoidal structure, and it will be nice to be able to construct arbitrary unitaries in the language, which will greatly enhance its expressive power from current Clifford circuits to a universal language.

\paragraph{Mechanization} There are multiple implementations of HTT in Coq such as Ynot~\parencite{ynot2008}. We would like to mechanize QHTT in Coq, F*, or some such dependent type theory for higher assurance of the usefulness and soundness of QHTT. This will also enable us to extract verified circuits in a lower level quantum language such as OpenQASM~\parencite{cross2017} for execution on real machines.

\paragraph{Linearity} Peter Selinger and collaborators have recently proposed a linear dependent typed version of Proto-Quipper (dubbed Proto-Quipper-D)~\parencite{selinger2020,fu2020linear}. It is an interesting challenge to reconcile linearity in our theory based on their proposal.

\paragraph{Circuits as Arrows} Further, Proto-Quipper treats quantum circuits as first-class citizens of the language. We would like to explore modifying our theory to treat Quantum Hoare types as arrows instead of as monads as was suggested by \textcite{so-arrows}. It makes sense from the perspective of sequential composition as arrows can have an arbitrary number of inputs/outputs as opposed to monads.

\paragraph{Behavioral Types} Another avenue for exploration is to incorporate more precise types for qubits that can distinguish between qubits in the pure classical state vs. those in superposition vs. those in entanglement~\parencite{JorrandPerdrix2009,Honda2015} such as those inspired by the various quantum resource theories or the Heisenberg representation of quantum mechanics~\parencite{rssl2019,rssl20}. For example, in \citetitle{rssl20}, \textcite{rssl20} assign types \textbf{Z} and \textbf{X} to single qubits in computational and Hadamard basis states respectively. Then, the Hadamard operator can be assigned the intersection of two arrow types, $(\mathbf{X} \rightarrow \mathbf{Z}) \cap (\mathbf{Z} \rightarrow \mathbf{X})$, as it converts between the computational and Hadamard basis states. Further, in that system, we can use the $\times$ type constructor to assign the type $\mathbf{X} \times \mathbf{Z}$ to the separable product state that we see for the pair of qubits $(a, b)$, right before applying the CX gate in \cref{lst:bell} to obtain the first Bell state. These and other such precise types can make our type system even more expressive.

\paragraph{Classical Effects} Finally, it will be an interesting challenge to reconcile both classical and quantum effects together in a single system as many interesting quantum algorithms involve a mixture of quantum operations and classical processing. It will be important to reason about the complete behavior of an implementation (as opposed to only the quantum effects) to be able to treat it as a reusable module. We see promise in exploring this direction using the F* programming language, where there is already existing work that lets one define additional effects relatively easily~\parencite{dm4free2017}.

\section*{Acknowledgments}
I would like to thank John Reppy and Robert Rand for their guidance. I would also like to thank the anonymous reviewers of QPL '20, and the larger QPL community for their feedback on an earlier presentation of this work~\parencite{qhtt20}. Casey Duckering, Anne Farrell, Sophia Lin, and Jocelyn Wilcox read the manuscript and provided helpful feedback.

This material is based upon work supported by EPiQC,\footnote{EPiQC: Enabling Practical-scale Quantum Computation. {\scriptsize URL:} \url{https://epiqc.cs.uchicago.edu}} an NSF Expedition in Computing, under Grant No. 1730449. Any opinions, findings, and conclusions, or recommendations expressed in this material are those
of the author and do not necessarily reflect the views of the National Science Foundation.

\printbibliography[heading=bibintoc]

\appendix

\newpage

\section{Grammar}
\label{app:grammar}

\nonterms{Typ}
\nonterms{Prop}
\nonterms{Tm}
\nonterms{Cmd}
\nonterms{Unitary}
\nonterms{Comp}
\nonterms{Ctx}

\section{Statics}
\label{app:rules}

\subsection{Valid types}

\drules[ty]{$\Gamma \vdash A$ \textbf{type}}{type formation}{Unit,Bit,Qbit,Prop,Pi,Pair,Hoare}

\subsection{Terms}

\drules[tm]{$\Gamma \vdash M : A$}{term typing}{VarE,UnitI,ZeroI,OneI,PiI,PiE,PairI,FstE,SndE}

Most of these rules are standard, but the following rule for suspended computation, which introduces a Hoare type, requires a bit of explanation:
\begin{mathpar}
    \drule{tm-HoareI}
\end{mathpar}

Here, the operator $*$ is analogous to a similar operator (\textit{separating conjunction}) in Separation Logic that says that if $R * S$, then $R$ and $S$ hold of disjoint portions of the state. So, in the assumption of the premise, $P * \top$ is a large footprint proposition that separates the portion of the state that the computation $E$ requires from the rest of the state using the trivial $\top$ predicate (which signifies that the computation $E$ knows nothing about the rest of the state). Further, the difference operator, $P \multimap Q$, says that the portion of the state satisfied by $P$ is replaced by a portion of the state that satisfies $Q$. In other words, the computation $E$ does not affect the rest of the state. This frame rule is essential to be able to provide small footprint style specifications in HTT, and hence, in our system. We leave a precise definition of the quantum analogs of separating conjunction and difference operator to future work.

\subsection{Computation rules}

\drules[comp]{$\Gamma, P \vdash E :  (x:A.\{Q\})$}{computations}{Consq,Return}

Notice that the \rref{comp-Consq} requires that we show that the postcondition $Q$ implies $R$ (denoted by $\models$ that denotes semantic consequence) in the appropriate context. Like the introduction \rref{tm-HoareI} for the Hoare type, its elimination rule also requires some explanation:
\begin{mathpar}
    \drule{comp-HoareE}
\end{mathpar}

Here, $K$ is a suspended computation (with a Hoare type) sequenced with another computation $E$. We need to show that the overall precondition, $P$, implies the large footprint proposition, $Q_1 * \top$, i.e., a portion of state satisfying $Q_1$ exists as required by $K$. The last premise says that in the expanded context, after executing $K$, computation $E$ is well-typed.

\subsection{Rules for primitive quantum commands}
We showed the rules for primitive commands in the form of quantum Hoare types in \cref{sec:primcmds}, here we show them in a natural deduction style for completeness. The main thing to notice about these rules is that they specify the strongest postcondition that captures the most precise relationship between the initial state and the modified state after the command's execution.

\begin{mathpar}
    \drule{cmd-Init}
\end{mathpar}
where $x$ is a fresh qubit variable and $\psi$ is either $\ket{0}$ or $\ket{1}$ depending on the input bit.

\begin{mathpar}
    \drule{cmd-Meas}
\end{mathpar}
where $x$ and $e_x$ are fresh variables and the strongest postcondition marks that the measured qubit is now in a classical state and a ghost qubit, $e_x$, now holds the state, $\psi$, previously held by the measured qubit.

Lastly, we show the rule for a single-qubit gate application:
\begin{mathpar}
    \drule{cmd-Unitary}
\end{mathpar}
where $U \cdot P$ corresponds to the unitary application to the overall state by appropriate padding.

\end{document}